\newcommand{\be}{\begin{equation}}
\newcommand{\ee}{\end{equation}}

\newcommand{\sign}{\mathop{\rm sign}\nolimits}
\newcommand{\tr}{\mathop{\rm tr}\nolimits}

\documentstyle[aps,epsf,multicol]{revtex}
\begin{document}
\draft
\widetext
\title{The $\theta$ vacuum reveals itself as the fundamental theory of the 
quantum Hall effect ~~~~~~~~~~~~~~~~II. The Coulomb interaction}
\author{A.M.M. Pruisken$^{1,2}$, M.A. Baranov$^{2,3}$ and 
I.S. Burmistrov$^{4}$ }

\address{$^{1}$ Indian Institute of Science, Bangalore 560012, India}
\address{$^{2}$ Institute for Theoretical Physics, Valckenierstraat 65, 1018XE Amsterdam, The Netherlands}
\address{$^{3}$ RRC Kurchatov Institute, Kuchatov Square 1, 123182 Moscow, Russia}
\address{$^{4}$L.D. Landau Institute for Theoretical Physics, Kosygina str. 2, 117940 Moscow, Russia }

\maketitle

\begin{abstract}
\noindent{Within the Grassmannian $U(2N) / U(N) \times U(N)$ non-linear 
$\sigma $ model representation of localization one can study the low energy 
dynamics of both the free and interacting electron gas. We study the 
cross-over between these two fundamentally different physical problems.
We show how the topological arguments for the exact quantization of the Hall conductance
are extended to include the Coulomb interaction problem. We discuss dynamical scaling 
and make contact with the theory of variable range hopping.}
\end{abstract}


\begin{multicols}{2}
\narrowtext

\begin{figure}
\begin{center}
\setlength{\unitlength}{1mm}
\begin{picture}(70,70)(0,0)
\put(0,0)
{\epsfxsize=70mm{\epsffile{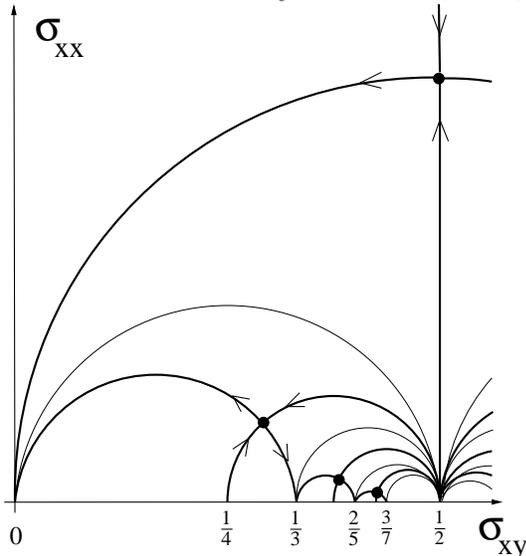}}}
\end{picture}
\caption{Unified scaling diagram for the quantum Hall effect
in the $\sigma_{xx}$, $\sigma_{xy}$ conductivity plane. 
The arrows indicate the direction towards the infrared.}
\label{mappedRG}
\end{center}
\end{figure}

Over the last few years, much effort has been devoted
to the problem of localization and interaction effects
in the quantum Hall regime~\cite{EuroLett,paperI,paperII,paperIII}. By now it is well understood that the
Coulomb interaction problem falls in a non-Fermi liquid 
universality class of transport problems with a novel
symmetry, named $\cal F$ {\em invariance}~\cite{paperI}. Although the
results for scaling are in many ways similar to those obtained for 
the free electron gas~\cite{freepart}, 
it is important to bear in mind that the Coulomb interaction problem 
is a far richer one. 
Unlike the free particle problem, for example, the field theory for
interacting particles provides the platform for a unification of
the {\em fractional quantum Hall regime}
and the quantum theory of {\em metals}~\cite{paperI,paperIII,fracedge}.
The principal features of this unification are encapsulated in a scaling 
diagram for the {\em longitudinal} and {\em Hall conductances} $\sigma_{xx}$ 
and $\sigma_{xy}$ respectivily (Fig. \ref{mappedRG}). The
\noindent{Finkelstein} approach to localization and
interaction phenomena~\cite{Finkelstein,italians}, the topological concept 
of an instanton vacuum~\cite{freepart}
as well as the Chern Simons statistical gauge fields~\cite{cf} are all essential
in composing this diagram.\\
\indent{The} main objective of this Letter is to embark on the most fundamental
aspect of the theory, the {\em observability} and {\em precision} 
of the quantum Hall effect. This experimental phenomenon is represented 
in Fig. \ref{mappedRG}  by the infrared stable fixed points located at
$\sigma_{xx} =0$ and $\sigma_{xy} = k$ (integer effect) as well as
$\sigma_{xy} = k/(2k+1)$ (Jain series). These fixed points, however, 
define the strong coupling phase of the unifying action where 
analytic work is generally impossible. In spite of ample experimental
evidence for its existence, 
the {\em robust quantization} of the Hall conductance has yet 
to be established as a fundamental but previously unrecognized feature 
of the topological $\theta$ vacuum concept.\\
\indent{In} what follows we shall benefit from the advancements reported in the preceding 
Letter. In particular, since the Finkelstein theory is formally defined 
as a $\sigma$ model
on the Grassmann manifold ${U(2N)}/{U(N) \times U(N)}$ 
with $ N$ equal to $N_r$ (number of {\em replica's}) 
times $N_m$ (number of {\em Matsubara frequencies}), 
we can now use our general knowledge on the strong coupling behavior 
of the theory and probe, for the first time, the quantum Hall phases
in the interacting electron gas.\\
\indent{To} achieve our goals we first shall outline 
some of the recent advancements in the field.
It is important to emphasize that the {\em complete} effective action for 
interacting particles now exists~\cite{paperI}. 
This action includes the coupling to external potentials and/or Chern Simon 
gauge fields.
This leads to a detailed understanding
of the electrodynamic $U(1)$ gauge invariance and provides invaluable
information on the renormalization of the theory that was not available 
before.
\indent{Secondly,} it is necessary to have a more detailed understanding of 
how the subtleties of interaction effects can be understood as a field theory.
For this purpose we report new results on the Grassmannian non-linear $\sigma $ model
with $N_r =0$ and varying $N_m$. These show explicitly how the cross-over takes
place between a theory of {\em free particles} at finite values of $N_m$ and a {\em many body} 
theory that is generally obtained in the limit $N_m \rightarrow \infty$ only. 
Armed with these insights we next point out how the Coulomb interaction problem, at
zero temperature ($T$), displays the general topological features and 
$\theta$ dependence that were discovered in the preceding Letter.\\
\indent{As} a third and final step toward the strong coupling phase we 
discuss the subject of dynamical scaling.
As a unique product of our effective action procedure we obtain a distinctly different 
behavior at finite $T$, depending on the specific regime of the interacting electron gas
that one is interested in. We establish, at the same time, the contact with the theory for 
{\em variable range hopping}~\cite{hopping}.
 
\vskip  1mm 
\noindent{$\bullet$ } {\em The action.~~~}
Following Finkelstein~\cite{Finkelstein}, the effective quantum theory for 
disordered (spin polarized or spinless) electrons is given in terms 
of a generalized $\sigma$ model involving the unitary matrix field variables 
${Q}_{n,m}^{\alpha,\beta} (\vec{x} )$ 
which obey $Q^2 = {\bf 1}$. Here, 
$\alpha, \beta$ represent the replica indices, $n,m$ are the indices of the Matsubara frequencies
$\nu_{k} = \pi T(2k+1)$ with $k=n,m$. In terms of ordinary unitary rotations 
${\cal V}_{n,m}^{\alpha,\beta}$ one can write

\be
Q={\cal V}^{-1} \Lambda {\cal V},~~~\Lambda =\Lambda_{n,m}^{\alpha,\beta} = \sign (\omega_n ) 
\delta_{n,m}^{\alpha,\beta},
\label{Qmatrix}
\ee
indicating that the $Q$ describes a Goldstone manifold
of a broken symmetry between positive and negative frequencies. 
A $U(1)$ gauge transformation in frequency space is
represented by a unitairy matrix ${\cal W}_{n,m}^{\alpha,\beta}$

\be
{\cal W} = \exp \{i \sum_{n , \alpha} \phi^\alpha (\omega_n ) {I}_n^\alpha \} .
\label{Wmatrix}
\ee
Here,  
$[{I}_k^\gamma ]_{n,m}^{\alpha,\beta} = \delta^{\alpha\gamma} \delta^{\beta\gamma}\delta_{n,m+k}$ 
denote the $U(1)$ generators. 
In finite frequency space with a cut-off ($N_m$), the $I$ matrices no longer span a $U(1)$
algebra. To define the $U(1)$ gauge invariance in a truncated frequency
space we have developed a set of rules 
($\cal F$ {\em algebra}~\cite{paperI}). These involve
one more (frequency) matrix, 
$\eta_{n,m}^{\alpha,\beta} = n \delta_{n,m}^{\alpha,\beta}$, that is used to represent
$\omega_n$. 
The effective action for electrons in a static magnetic field $B$ and coupled to external 
potentials and/or Chern Simons fields $a_{\mu}^{\alpha}(\vec x , \omega_n)$
with $\omega_n = 2 \pi T n \neq 0$, 
can now be written as~\cite{paperI}

\be
	S_{eff} =  S_\sigma +S_F +S_U + S_0 .
\label{Seff}
\ee
Here, $S_{\sigma}$ is the free electron piece~\cite{freepart} 
\be
	S_{\sigma} = \frac{\sigma_{xx}^0}{8} \int {\rm tr} [\vec D,Q]^2
	+\frac{\sigma_{xy}^0}{8} \int {\rm tr} \epsilon_{ij} 
  Q[D_i,Q][D_j,Q] ,
\label{Ssigma}
\ee
where 
$D_j = \nabla_j -i \sum_{n\alpha} a_j^{\alpha} (\vec{x} , \omega_n ) {I}_n^{\alpha} $
is the covariant derivative.
Next, the two pieces $S_F$ and $S_U$ are linear in $T$ and represent
interaction terms. $S_F$ is gauge invariant and contains the 
{\em singlet interaction} term~\cite{Finkelstein}

\be
	S_{ F} = z_0 \pi T 
	\int \left[ {\sum_{\alpha n}} {\rm tr} {I}_n^\alpha Q
	{\rm tr} {I}_{-n}^\alpha Q +4 {\rm tr} \eta  Q - 6 {\rm tr} \eta  \Lambda
	\right].
\label{Sf}
\ee
The (Coulomb) term $S_U$ contains the scalar potential
$$
	S_{U} = -\pi T {\sum_{\alpha n}}\int_{x} \int_{x'} \left[
	\tr {I}_{-n}^\alpha Q(\vec{x})-\frac{1}{\pi T} {\tilde a}_0^\alpha (\vec{x}, \omega_{-n} )
	\right]
	\times
$$
\be
	\;\;\;\;\;\;\;\;\;\;\; U^{-1} (\vec{x} -\vec{x'} ) 	
	\left[
	\tr {I}_{n}^\alpha Q(\vec{x'})-\frac{1}{\pi T} {\tilde a}_0^\alpha (\vec{x'} , \omega_n )
	\right] .
	\label{Su}
\ee
$S_0$ contains the magnetic field 
${b}^{\alpha}=\nabla_x a^\alpha_y - \nabla_y a^\alpha_x$ 
\be
	S_0 = -\frac{\rho_B^2 }{2\rho T}  
\int_x \sum_{\alpha n} b^{\alpha}(\vec{x} , \omega_n) 
	b^{\alpha}(\vec{x} , \omega_{-n}) .
\label{S0}
\ee
We have defined (dropping the replica index $\alpha$ on $a_{\mu}$) 
\be
	\tilde{a}_0 =a_0 - i \rho_B  b / \rho \;\;\; ; \;\; U(q) 
= \rho^{-1} + U_0 (q).
\label{aU}
\ee
Here, the density of states		
$\rho = {\partial n}/{\partial \mu}$ and the quantity $\rho_B = {\partial n}/{\partial B} $
are thermodynamic quantities.
The statement of gauge invariance now means that the theory is invariant under the following
transformation
\be
	Q \rightarrow {\cal W}^{-1} Q {\cal W} \;\;\; ; \;\;\; a_\mu \rightarrow a_\mu + \partial_\mu \phi .
\label{gauge}
\ee
Using Eq. (\ref{gauge}) it is easy to see that the action is invariant under spatially
independent gauge transformations $\phi =\phi (\omega_n)$ provided the interaction 
potential $U_0$ 
has an infinite range. This global invariance, termed $\cal F$ {\em invariance}, 
is an exact symmetry of the Coulomb interaction problem which in two spatial
dimensions is represented by $U_0^{-1} (q) = {\Gamma |q|}$.

\vskip  1mm
\noindent{$\bullet$} {\em Static versus dynamic response.~~~}
Our introduction of external potentials (statistical gauge fields) 
$a_\mu$ can be exploited immediately to elucidate fundamental aspects of 
the quantum transport problem in strong $B$.
For this purpose we consider $S_{eff} (a_\mu)$ obtained after elimination 
of the $Q$ fields. 
Defining the particle density $n = T { \delta S_{eff} }/{\delta  a_{0} }$ we obtain,
at a tree level, the continuity equation 
$$
\omega_m (n+i\sigma _{xy}^{0} b ) = 
\vec{\nabla} \cdot \left[ \sigma _{xx}^{0} ( \vec{e} + \vec{\nabla} U_{0} n ) 
- D_{xx}^0 \vec{\nabla} ( n + i \rho_B b ) \right] .
$$
Here, $D_{xx}^0 = \sigma_{xx}^0/\rho $ denotes the diffusion constant and 
$\vec{e}$, $b$ are the external electric and magnetic fields respectively. 
This result is familiar from the 
theory of metals~\cite{NozPines} where the quantity $\rho_B$ is usually neglected.
Notice that in the static limit $\omega_n \rightarrow 0$ both quantities $\sigma_{ij}^0$ 
drop out and the equation now contains the thermodynamic 
quantities $\rho, \rho_B$ and $U_0$ only. Since the fields
$a_\mu (\vec{x} , \omega_n =0)$ are completely
decoupled from the $Q$ field variables, the static response
is actually determined by a different, underlying theory.
This means that 
$\rho, \rho_B$, $U_0$ and hence $S_U$ and $S_0$ 
should not have any quantum corrections in general, either perturbatively or
non-perturbatively. This observation can be used as a 
general physical constraint that must be imposed on the quantum theory.
The only quantities that are allowed to have quantum corrections are the transport 
parameters $\sigma_{xx}^0$, $\sigma_{xy}^0$, and the singlet interaction amplitude $z_0$. 

As an important check on the statements of gauge invariance and renormalization,
we have evaluated the quantum theory in $2+\epsilon$ spatial dimensions to order 
$\epsilon^2$. 
The results of the computation, along with an extensive analysis of dynamical scaling, 
have been reported in~\cite{twoloop}. 

\vskip  1mm
\noindent{$\bullet$} $\cal F$ {\em invariance.~~~} The
renormalizability can be addressed more formally, by making contact with the theory of 
ordinary non-linear sigma models~\cite{sigma}. 
For this purpose we drop the external potentials from
the action and recall that for finite size matrices $Q$, operators like $S_F$ 
play the role of infrared regulators that 
do not affect the singularity structure of the theory at short distances.
We know in particular that the theory is renormalizable in two 
dimensions. 
Besides the coupling constant or $\sigma_{xx}^0$, 
one additional renormalization constant is needed for the operators 
linear in the $Q$ matrix field and two more are generally needed for the operators 
bilinear in the $Q$ (i.e. the symmetric and anti-symmetric representation respectivily)~\cite{sigma1}.
These general statements apply to the Finkelstein action as well 
since the latter only demands that the number of Matsubara frequencies $N_m$ is taken to infinity 
(along with $N_r \rightarrow 0$). To completly undust this point we have computed the 
cross over functions for the theory where the quantity $U^{-1} ({\vec x} - {\vec x}' )$ in $S_U$, 
Eq.(\ref{Su}), is replaced by its most relevant part

\be
	U^{-1} ({\vec x} - {\vec x}' ) \rightarrow z_0 (1-c_0 ) \delta ({\vec x} - {\vec x}' ) .
\label{zc}
\ee
Notice that $0 < c_0 < 1$ represents the
finite range interaction case. The extreme cases $c_0 = 0$ and $1$ describe the free electron gas 
and the Coulomb system respectivily. $\cal F$ invariance is retained for $c_0 =1$ only
and broken otherwise.

The following renormalization group functions have been obtained for the parameters $z$, 
$c$ and the dimensionless resistance $g=\mu^\epsilon /\pi\sigma_{xx}$ in $2+\epsilon$ 
spatial dimensions ($\mu$ denotes an arbitrary momentum scale) 

\begin{eqnarray}
	\frac{dg}{d\ln\mu} & = & \epsilon g - 2 g^2 ( f + \frac{1-c}{c} \ln (1-cf)  ) 
	\nonumber \\
	\frac{d\ln z}{d\ln\mu} & = & g c f,\;\;\;\;\;\;\;
	\frac{d c}{d\ln\mu} =  g c( 1- cf).
\label{RGf}
\end{eqnarray}
Here, $f = M^2/(\mu^2 + M^2)$ is a $\mu$-dependent function with 
$M^2 = 8\pi z_0 T N_m /\sigma_{xx}^0$ which depends on the cut-off $N_m$.

For $f=0$ ($\mu >> M$ or short distances) we obtain the well known
results for free particles~\cite{freepart,sigma}, i.e. 
$dg/d\ln\mu$ has no one-loop contribution, 
$z$  has no quantum corrections in general and the result for $c$ 
coincides with the renormalization of symmetric operators, bilinear in $Q$. 

For $f=1$ ($\mu << M$ or large distances), we obtain 
the peculiar Finkelstein 
results of the interacting electron gas~\cite{paperII,italians}. 
The symmetry breaking parameter $c$ now affects the renormalization 
of all the other parameters. The concept of $\cal F$ invariance 
($c = 1$) manifests itself as
a new (non-Fermi liquid) fixed point in the theory. The problem 
with $0 < c < 1$ 
lies in the domain of attraction
of the Fermi liquid line $c =0$ which is stable in the infrared.

Notice, however, that the $\cal F$ invariant fixed point $c = 1$ 
only exists if the mass $M$ in the theory
remains finite at zero $T$. This clearly shows that, in order for 
$\cal F$ invariance
to represent an exact symmetry of the problem, $N_m$ must be infinite. 
The time $\tau$ plays the role of an extra, non-trivial 
dimension and this dramatically complicates the problem of plateau
transitions in the quantum Hall regime. The Coulomb interaction problem,
unlike the free electron theory, is given  
as a $2+1$ dimensional field theory, thus invalidating any attempt 
toward exact solutions of the experimentally observed critical 
indices~\cite{experiments}. 

\vskip  1mm
\noindent{$\bullet$} {\em The qHe.~~~}
Next, we turn to the most interesting aspect of the theory, the 
$\sigma_{xy}^0$ term ($\theta$ term), which is invisible in perturbative
expansions. However, we may proceed along the same lines as pointed out
in the previous Letter and separate, in the theory for $T=0$, 
the {\em bulk} quantities from the {\em edge} quantities~\cite{largeN}.
Specifying to the Coulomb interaction problem ($c=1$) in two spatial dimensions,
we next make use of the principle
of $\cal F$ invariance and formulate an effective action for the edge. In the notation
introduced before we now have
\begin{eqnarray}
	S_{eff} (q) & = & S_{bulk} (q) + 2\pi i k(\nu) C(q) , 
\nonumber \\
e^{-S_{bulk} (q)} & = & \int_{\partial V} D[Q_0 ] e^{-{\tilde S}_{\sigma} 
( t^{-1} Q_0 t) - S_F (t^{-1} Q_0 t)} .
\label{Sb}
\end{eqnarray}
Here ${\tilde S}_{\sigma}$ is the same as $S_\sigma$ with $\sigma_{xy}^0$ replaced
by its {\em unquantized} bulk piece $\theta(\nu)$. Recall that the 
functional integral is performed with a fixed value $Q_0 = \Lambda$ at the edge. 
It is important to notice that the interaction piece $S_F$ cannot be left out
since it affects, following Eq. (\ref{RGf}), the renormalization of the theory at $T=0$.

The definition of $S_{eff} (t)$ is precisely the same as the 
{\em background field} methodology adapted to the Coulomb interaction 
problem~\cite{paperII,EuroLett}. The result is of the form

\be
	{S_{bulk} (q)} = F(\theta) + {\tilde S}_{\sigma} ' (q) + S_F ' (q) ,
\label{Sb2}
\ee
where the primes indicate that the parameters $\sigma_{xx}^0$, $\theta(\nu)$ and $z_0$ are
replaced by renormalized ones, $\sigma'_{xx} $, $\theta' $ and $z'$ 
respectivily, which are defined for system size $L$. 

This leads to the most important statement of this Letter which says
that, provided a $mass$ is generated for {\em bulk} excitations, the
renormalized theory $\sigma'_{xx} =\sigma_{xx} (L)$, $\theta' =\theta (L)$ and $z'=z (L)$ 
should vanish for large enough $L$,
i.e. the bulk of the system is insensitive to changes in the boundary
conditions except for corrections exponentially small in $L$.
Under these circumstances $S_{eff}[q]$ reduces to the
action of {\em massless chiral edge excitations}~\cite{paperIII,fracedge}.
The integer $k(\nu)$ equals
the number of edge modes and is now identified as the 
{\em quantized Hall conductance}. 

These results describe the strong coupling 
``integer quantum Hall'' fixed points (Fig. \ref{mappedRG}) that were previously conjectured on 
phenomenological grounds. From the weak coupling side, a detailed analysis of Eqs. (\ref{Sb}), (\ref{Sb2}) 
leads to the following expressions~\cite{EuroLett} for the renormalization group functions 
$d\sigma_{xx}/d\ln\mu = \beta_\sigma (\sigma_{xx} , \theta)$,
$d\theta/d\ln\mu = \beta_\theta (\sigma_{xx} , \theta)$ and
$d\ln z/d\ln\mu = \gamma_z (\sigma_{xx} , \theta)$, 

\begin{eqnarray}
	\beta_\sigma & = & \beta_\sigma^0 (\sigma_{xx} )
	+ D_0 \sigma_{xx}^2 e^{-2\pi\sigma_{xx}} \cos \theta
	\nonumber \\ 
	\beta_\theta & = & ~~~~~~~~~~~~~~D_0 \sigma_{xx}^2 e^{-2\pi\sigma_{xx}} \sin \theta
	\nonumber \\
	\gamma_z & = & \gamma_z^0 (\sigma_{xx} ) 
	+ D_0 \sigma_{xx} e^{-2\pi\sigma_{xx}} \cos \theta,
\label{RGi}
\end{eqnarray}
Here, $D_0$ is a positive constant determined by the instanton determinant and
$\beta_\sigma^0 $ and $\gamma_z^0 $ are the perturbative results
that recently have been extended to two-loop order ($A \approx 1.64$)~\cite{paperII,twoloop} 
\be
	\beta_\sigma^0 (\sigma_{xx} )  = \frac{2}{\pi} + \frac{4A}{\sigma_{xx}};~~
	\gamma_z^0 (\sigma_{xx} )  =   \frac{1}{\pi\sigma_{xx}} + \frac{18+\pi^2}{6\pi^2\sigma_{xx}^2} .
\label{2loop}
\ee
In summary, there is now fundamental support, both from the weak and strong coupling side, 
for the scaling diagram of the integral quantum Hall effect~\cite{EuroLett}. 
\vskip  1mm
\noindent{$\bullet $} {\em Finite T.~~~} At finite $T$ the infrared of the system is
controlled by the interaction terms $S_F$ and $S_U$. In this case one must go back 
to the original theory (Eqs. (\ref{Seff})-(\ref{S0})) and obtain the transport parameters from linear 
response in the field $a_\mu$~\cite{paperII}.
Specifying to the $a_0 = 0$ gauge as well as $\vec{\nabla} \cdot \vec{a} = 
\vec{\nabla} \times \vec{a} = 0$ we can write
\begin{eqnarray}
S_{eff} (a_j ) = T\sum_{n>0} \int_x \omega_n \left[ \sigma'_{xx} \delta_{ij} + 
\sigma'_{xy} \epsilon_{ij} \right] a_i (\omega_n ) a_j (-\omega_n ) 
\nonumber
\end{eqnarray}
where the expressions for $\sigma'_{ij}$ are known as the {\em Kubo formulae}~\cite{paperII}. 
We stress that these expressions are
exactly the same as those obtained from the background field proceedure, 
$\sigma'_{xx} =\sigma_{xx}(L)$ 
and $\sigma'_{xy} = k(\nu) + \theta (L)/2\pi$ (Eqs \ref{Sb2}, \ref{RGi}),
provided $S_{eff} (a_j ) $ is evaluated at $T=0 $ and with $Q=\Lambda$ at the 
edge~\cite{paperI,paperII}

The scaling results at finite $T$ generally depend on the 
specific regime and/or microscopics of the disordered electron gas that one is
interested in. Here we consider the most interesting cases where $\theta (\nu) \sim  \pm \pi $ 
and $\theta (\nu) \sim 0$ respectivily. The first case is realized when
the Fermi level passes through the center of the Landau 
band where the electron gas is quantum critical and the transition takes place between
adjacent quantum Hall plateaus~\cite{freepart}. 
Provided the bare parameter $\sigma_{xx}^0$
of the theory is close to the critical fixed point $\sigma_{xx}^* $ at $\sigma_{xy} =1/2$
(Fig. \ref{mappedRG})
the following universal scaling law is observed~\cite{freepart}

\begin{equation}
\sigma'_{xx} = \sigma_{xx} (X) ~~~ \sigma'_{xy} = k(\nu) + {\theta (X)}/{2\pi} 
\label{ss}
\end{equation}
where $X = (zT)^{-\kappa } \Delta \nu$. Here $\Delta \nu$ is the filling fraction $\nu$ of the
Landau levels relative to the critical value $\nu^*$ which is half-integer. 
The correlation (localization) length $\xi $ of the electron gas
diverges algebraically $\xi \propto |\Delta\nu |^{-1/y_\theta}$. The critical indices $\kappa$ 
and $y_\theta$ are a major objective of experimental research~\cite{experiments} and 
the results have been discussed extensively and at many places~\cite{EuroLett,paperIII,twoloop,hopping}.  

Next we consider $\theta(\nu) \sim 0$ which is entirely different. This happens when 
the Fermi energy is located at the tail end of the Landau bands corresponding to the 
center of the quantum Hall plateau. 
The bare parameter $\sigma_{xx}^0$ of the theory is now close to zero~\cite{paperIII}. 
This means that the $T$ dependence is determined by the strong coupling 
asymptotics of the renormalization ($\theta ,~\sigma_{xx} \rightarrow 0$). Notice
that the $\gamma_z$ function (Eqs. \ref{RGi}, \ref{2loop}) indicates that the singlet interaction term 
$S_F$ eventually renders irrelevant as compared to 
the Coulomb term $S_U$ (with $U^{-1} (q) =\Gamma |q|$) which, as we mentioned before, 
is not affected by the quantum theory. One now expects $S_U$ to become the dominant
infrared regulator such that the scaling variable $X$ in Eq. (\ref{ss}) is now given by
$X = T\Gamma \xi$. 

This asymptotic limit of the theory 
can be identified as the Effros - Shklovskii regime of 
{\em variable range hopping} for which the following result is known 
$\sigma'_{xx} = \sigma_{xx} (T\Gamma \xi ) = \exp \{ -2/ \sqrt{T\Gamma \xi } \}$~\cite{hopping}.
We therefore conclude that the dynamics of the electron gas is generally described 
by distinctly different physical processes and controled by completely different fixed points 
in the theory.

The research was supported in part by FOM and INTAS 
(Grant 99-1070).

\end{multicols}
\end{document}